\documentclass[aps,pre,reprint,groupedaddress,showpacs,showkeys]{revtex4-1}
\usepackage{graphicx}

\begin{document}

\title{Poresizes in random line networks}

\author{Claus Metzner}
\email[E-mail me at: ]{claus.metzner@gmx.net}
\author{Patrick Krauss}
\author{Ben Fabry}
\affiliation{Biophysics Group, University of Erlangen, Henkestr. 91, D-91052 Erlangen, Germany}

\date{\today}

\begin{abstract}
Many natural fibrous networks with fiber diameters much smaller than the average poresize can be described as three-dimensional (3D) random line networks. We consider here a `Mikado' model for such systems, consisting of straight line segments of equal length, distributed homogeneously and isotropically in space. First, we derive analytically the probability density distribution $p(r_{no})$ for the `nearest obstacle distance' $r_{no}$ between a randomly chosen test point within the network pores and its closest neighboring point on a line segment. Second, we show that in the limit where the line segments are much longer than the typical pore size, $p(r_{no})$ becomes a Rayleigh distribution. The single parameter $\sigma$ of this Rayleigh distribution represents the most probable nearest obstacle distance and can be expressed in terms of the total line length per unit volume. Finally, we show by numerical simulations that $\sigma$ differs only by a constant factor from the intuitive notion of average `pore size', defined by finding the maximum sphere that fits into each pore and then averaging over the radii of these spheres. 
\end{abstract}



\maketitle

\section{Introduction}

Many biological systems can be structurally described as random line networks. A typical example are gels that self-organize by the polymerization and subsequent crosslinking of filamentous proteins, such as collagen or fibrin. In order to characterize the stochastic geometry of such systems, a frequently used parameter is the average poresize $\overline{r}_{pore}$. It is determined by finding, for a reprensentative fraction of network pores, the largest spheres that can be fit into that pores and then computing the average of the radii of these maximum spheres. While this can be done numerically in a straight forward yet time consuming way, this definition of poresize is not suited very well for exact analytical calculations. Therefore, we suggest as an alternative measure the most probable nearest obstacle distance $\sigma$ for randomly chosen test points and show that it is directly related to the poresize. 

\begin{figure}
\includegraphics[width=0.65\linewidth]{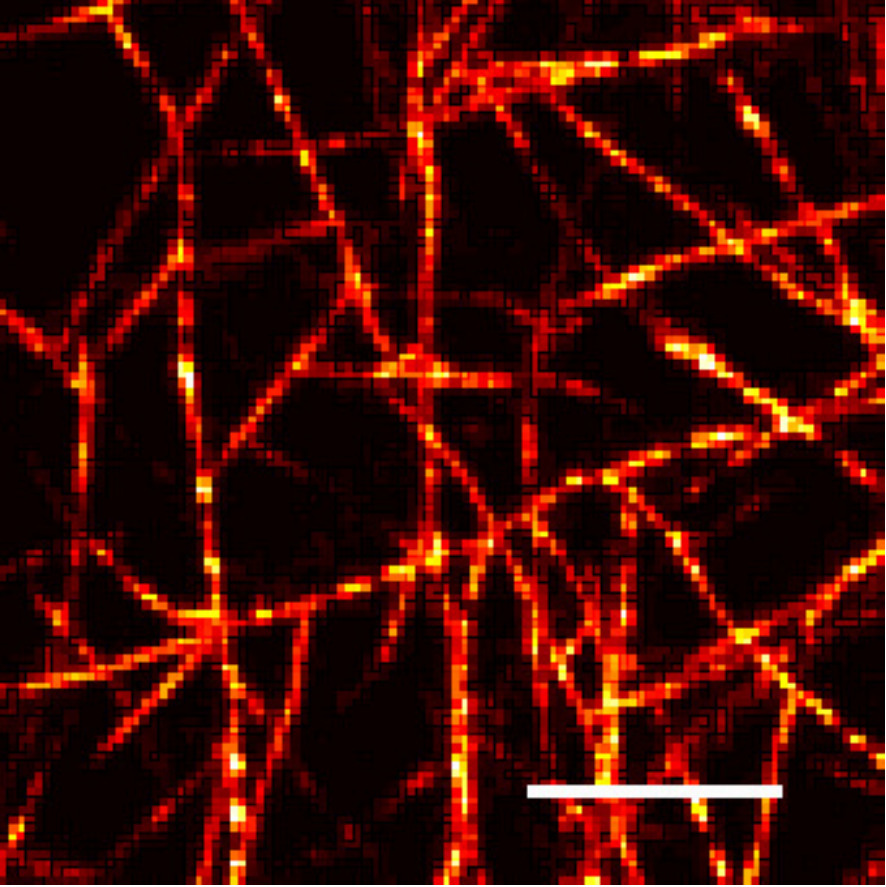}
\caption{\label{fig:fig0} Collagen gel with a concentration of 1.2 mg/ml (Scale bar is 10 $\mu$m). Shown is the maximum intensity projection from a stack of 15 single confocal images recorded at a z-distance of 340 nm (total height = 5.1 $\mu$m). The fibers are straight on the scale of a typical pore size $\overline{r}_{pore}$ and their diameter is much smaller than $\overline{r}_{pore}$. Therefore, the structure of the system can be well approximated by a Mikado line network in the long fiber limit.}
\end{figure}

We start our investigations with a `Mikado'-like network model that has two parameters, the length $l$ of the line segments and the volume density $\rho$ of their centers. It is possible to compute the distribution $p(r_{no})$ of nearest obstacle distances analytically in this model. In the limit of zero line length, the Mikado model contains the case of point networks. More interesting is the opposite limit, where $l$ is much larger than the average poresize. In this case, the Mikado model converges towards a more general model that represents any random line network with a large persistence length. Indeed, the single system parameter in this limiting case is the overall line density $\lambda$, i.e. the total line length per unit volume. This parameter only sets the spatial scale of the network, and no other details matter for the distributions $p(r_{no})$ or $W(r_{pore})$. For example, a network composed of random circles with identical overall line density would yield the same universal Rayleigh distribution $p(r_{no})$ as the Mikado model, provided the radius of the circles is much larger than the average pore size. We compare these analytical results to a numerical simulation that is directly based on the exaxt analytic geometry of points and lines, thus avoiding any possible artifacts arising from voxelation.

After demonstrating perfect agreement of the simulations with the analytic results, we use the simulations to determine the poresize distribution for line networks of various density parameters $\lambda$. As expected from scaling arguments, the average poresize $\overline{r}_{pore} = c \sigma$ is simply proportional to $\sigma$, allowing us to determine the conversion factor as $c\approx$1.86.

\section{Model and Theory}

\subsection*{Distribution of nearest obstacles distances $p(r_{\rm{no}})$, accessible volume fraction $Q(r)$ and pore sizes}
\begin{figure}
\includegraphics[width=0.95\linewidth]{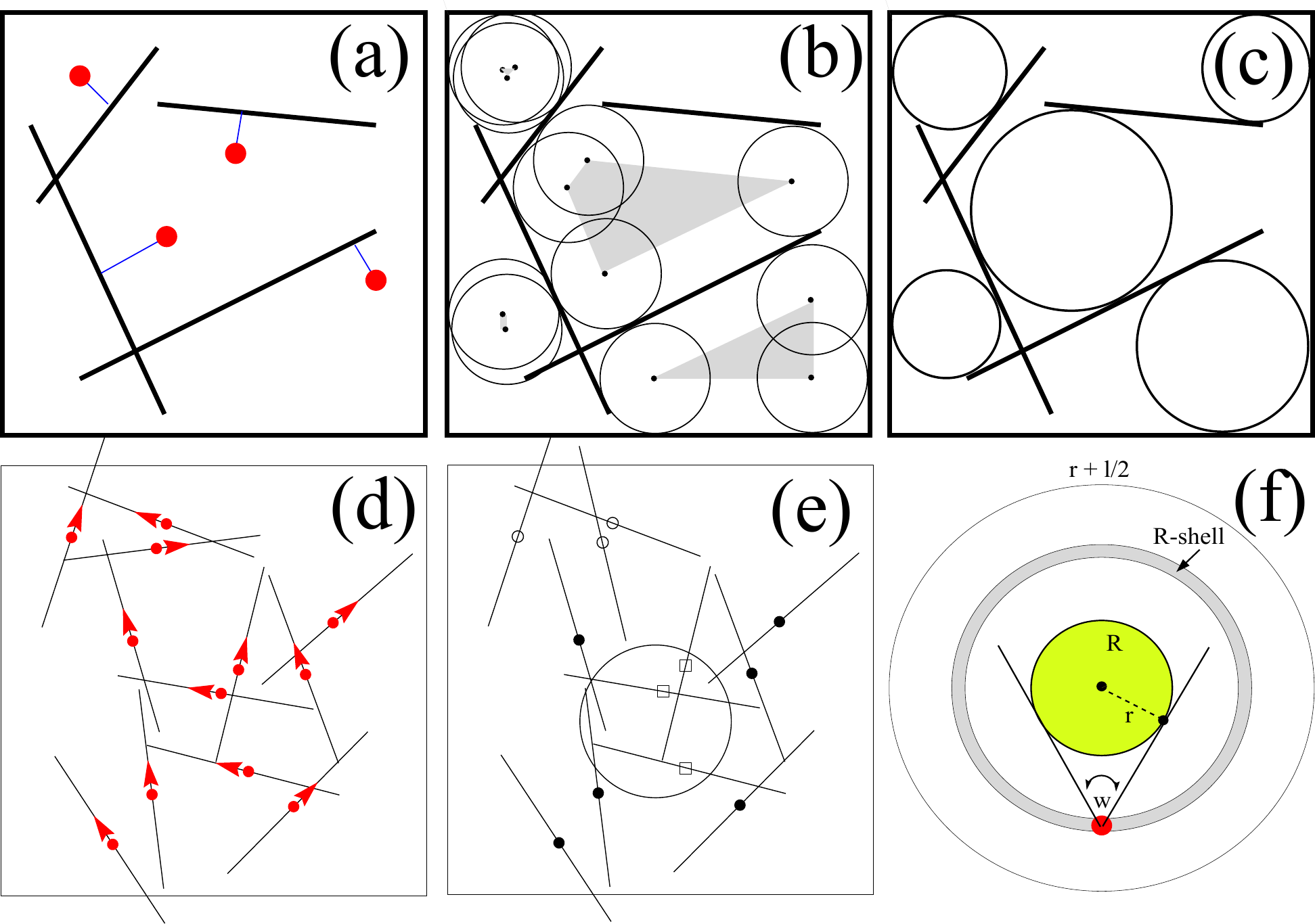}
\caption{\label{fig:fig1} \textbf{2D illustration of various statistical measures used for networks of line segments.} (a) Nearest obstacle distances $r_{\rm{no}}$ (thin lines) for a few selected points (circles). (b) Accessible volume (shaded areas) for spheres of a given radius. (c) Maximum spheres fitting into network pores, thereby defining the pore sizes $r_{pore}$. (d) A homogeneous, isotropic random distribution of straight line segments. The segments have a prescribed length $l$ and their center points a spatial density of $\rho$. (e) Classification of line segments in the 2D Mikado model. 1-group (squares): Centers within r-sphere. 2-group (full circles): Centers outside r-sphere, yet with chance of overlap. 3-group (empty circles): Remote segments without chance of overlap. (f) 2D sketch of an r-sphere (green), a concentric spherical shell of radius $R$ (gray) and a specific point (red) within this shell. From all line segments centered at the red point, only those can intersect the r-sphere with orientations falling into a cone of apex angle $\omega$.}
\end{figure}

We consider random biphasic networks, in which every point of 3-dimensional space either belongs to phase 0 (pore, liquid) or phase 1 (material, solid). In order to map out the stochastic geometry of the network, one can repeatedly choose a random point $\vec{R}_0=(x,y,z)$ within the 0-phase of the network and then find its `nearest obstacle distance' $r_{\rm{no}}(\vec{R}_0)$, defined as the Euclidean distance from that point $\vec{R}_0$ to the closest point of the 1-phase (compare Fig. \ref{fig:fig1}(a)). The network is then characterized by the distribution $p(r_{\rm{no}})$ of the nearest obstacle distances.

Closely related to $p(r_{\rm{no}})$ is the `accessible volume fraction' $Q(r)$, defined as the fraction of the 0-phase in which a sphere of radius $r$ (from now on called a r-sphere) could be centered without overlapping the 1-phase (compare Fig. \ref{fig:fig1}(b)). In general, the dimensionless quantity $Q(r)$ has the value $Q(r\!=\!0)\!=\!1$ and decreases monotonically for all radii $r\!>\!0$.

The complemental quantity $1-Q(r)$ is the fraction of 0-phase for which an r-sphere overlaps the 1-phase. It corresponds to the probability that a random 0-phase point $\vec{R}_0$ has a nearest obstacle distance $r_{\rm{no}}$ smaller than $r$, or
\begin{eqnarray}
1-Q(r) &=& \mbox{Prob}(r_{\rm{no}}<r)\nonumber\\
&=& \int_0^r p(r_{\rm{no}}) dr_{\rm{no}}.
\end{eqnarray} 
The derivative of this equation with respect to $r$ shows that $Q(r)$ is just the negative cumulative probability of $p(r_{\rm{no}})$:
\begin{equation}
p(r=r_{\rm{no}}) = - \frac{d}{dr} Q(r).
\end{equation}
While both quantities carry the same information about the network, the cumulative $Q(r)$ is more convenient for analytical considerations, as will be demonstrated below.

Another way to characterize pores of a network is to find the maximum sphere that fits to each pore and to define the `pore size' $r_{\rm{pore}}$ as the radius of this maximum sphere. The concept is also illustrated in Fig. \ref{fig:fig1}(c). We denote the distribution of pore sizes by $W(r_{\rm{pore}})$.

\subsection*{Random Line Networks: The Mikado model}
In the following we consider random networks in which the 1-phase consists of straight line segments of fixed length, with isotropic orientations and a homogeneous distribution throughout the 3D volume. We refer to this model as the Mikado model.

Each individual line segment (LS) can be described by its center point and a unit direction vector. In order to avoid ambiguities, we require that all unit vectors have a positive z-component and thus `point upwards' (compare Fig. \ref{fig:fig1}(d)). The two parameters of the Mikado model are the length $l$ of the LSs and the volume density $\rho\!=\!\frac{N}{V}$ of their center points, where $N$ is the number of line segments within a volume $V$.

Consider first the extreme case $l\rightarrow0$, where all LSs degenerate into their center points, and place a r-sphere randomly into the system. Note that the configuration of LS-centers throughout the volume is a spatial Poisson process with `event rate' that is identical to the volume density $\rho$. On average, the r-sphere will contain a number of 
\begin{equation}
n_{av,l\rightarrow0}(r)=\rho\frac{4}{3}\pi r^3
\end{equation}
LS-centers. The probability $Q(r)$ that not a single LS-center lies within the r-sphere is given by the Poisson probability for $k\!=\!0$ events, which is
\begin{equation}
Q(r)=\mbox{Poisson}\left\{k=0,n_{av}=n_{av,l\rightarrow0}(r)\right\}=e^{-n_{av,l\rightarrow0}(r)}.
\end{equation}
Therefore, in the case of the random point network the accessible volume fraction is given by
\begin{equation}
Q(r)_{l\rightarrow0}=e^{-\frac{4\pi}{3}\rho r^3}.
\end{equation}
We now turn back to the general case $l>0$. As before, we can write
\begin{equation}
Q(r)=e^{-n_{av}(r)}.
\end{equation}

In order to compute $n_{av}(r)$, we note that with respect to a given r-sphere, the LSs can be classified into 3 groups (compare Fig. \ref{fig:fig1} (e)): 
\begin{itemize}
\item 1-group with LS-centers inside the r-sphere.
\item 2-group with LS-centers outside the r-sphere, but yet with a possibility of intersecting the r-sphere.
\item 3-group with LS-centers too far away to touch the r-sphere.
\end{itemize}
Only the groups 1 and 2 contribute to $n_{\rm{av}}(r)$. The contribution of the 1-group is identical to the case of point networks above:
\begin{equation}
n_{av,1}(r)=\rho\frac{4}{3}\pi r^3.
\end{equation}

The 2-group consists of LSs with centers in a sphere of radius $r+(l/2)$ around the center of the r-sphere. 
We now consider in more detail the ones in an infinitesimal spherical shell of radius $R$ around the center of the r-sphere, with $0\!<\!R\!<\!r\!+\!(l/2)$. This
R-shell contains a number of
\begin{equation}
dN^{\prime}=\rho dV=\rho 4\pi R^2dR
\end{equation}
candidates for intersection. Among them, only those LSs will actually overlap the r-sphere that have orientations within a certain cone (compare Fig. \ref{fig:fig1}(f)). This cone has an apex angle of $\omega=2\arcsin(r/R)$ and the corresponding solid angle is 
\begin{eqnarray}
\Omega(R)&=&4\pi\sin^2(\omega/4)\nonumber\\
&=& 4\pi\left[ \sin\left(\frac{1}{2}\arcsin(r/R)\right)\right]^2\nonumber\\
&=&2\pi\left(1-\sqrt{1-(r/R)^2}\right).
\end{eqnarray}
Since the total solid angle available for LS orientations is $\Omega_{tot}=2\pi$ (according to our convention that all unit direction vectors are pointing upward), the intersecting LSs amount to a fraction of $\Omega(R)/\Omega_{tot}=\left(1-\sqrt{1-(r/R)^2}\right)$. We conclude that the average number of actual intersections from LSs within the R-shell is
\begin{eqnarray}
dN(R)&=&dN^{\prime}\left(1-\sqrt{1-(r/R)^2}\right)\nonumber\\
&=&4\pi\rho \left(1-\sqrt{1-(r/R)^2}\right) R^2 dR.
\end{eqnarray}
The total contribution from all LSs of the 2-group is obtained by integration over the relevant R-shells:
\begin{equation}
n_{av,2}(r)=\int_{R=r}^{R=r+(l/2)} dN(R).
\end{equation}
This integral can be performed analytically. Using the abbreviation
\begin{equation}
f(s):= \frac{1}{3}\left[ s^3 - (s^2-1)^{3/2} \right],
\end{equation}
one obtains
\begin{equation}
n_{av,2}(r)= 4\pi\rho r^3 \left[ f(1+\frac{l}{2r}) - f(1) \right].
\end{equation}
By adding the contributions of both relevant groups, $n_{\rm{av}}(r)=n_{av,1}(r)+n_{av,2}(r)$, and using $Q(r)=e^{-n_{av}(r)}$, we arrive at an analytic expression for the accessible volume fraction in the Mikado model. Defining another useful abbreviation
\begin{equation}
g(x) := 3 \left[ f(1+\frac{x}{2})-\frac{1}{3}\right],
\end{equation}
the result can be cast into the form
\begin{equation}
Q(r)= e^{-\frac{4\pi}{3}\rho r^3\left[ 1 + g(l/r)\right]}.
\end{equation}
It correctly contains the limit of point networks, since $g(l/r)\rightarrow 0$ for $l \rightarrow 0$. All the differences between point and LS networks are included in the `perturbation function' $g(l/r)$. 

From the accessible volume fraction $Q(r)$, we immediately obtain the distribution of nearest obstacle distances $p(r_{\rm{no}}\!=\!r) = - \frac{d}{dr} Q(r)$ in the Mikado model. With increasing $r_{\rm{no}}$, this distribution starts with $p(r_{\rm{no}}\!=\!0)\!=\!0$, develops a single peak and then decays exponentially for distances much larger than the average pore size $\overline{R}_{pore}$ of the network. 

\subsection*{Mikado model in the long fiber limit}
We next consider the case $l \gg r$, where the LSs are much longer than the typical distances of interest. Since $p(r_{\rm{no}})$ is exponentially small for distances beyond the average pore size, this limit can also be interpreted as $l \gg \overline{R}_{\rm{pore}}$. Note that this is a typical situation for networks of semi-flexible fibers, such as collagen.

It is straight-forward to show that in this limit the perturbation function diverges as $g(l/r) \rightarrow \frac{3}{4}\frac{l}{r}$. One therefore obtains
\begin{equation}
Q_{l\gg r}(r)= e^{-(\pi\rho l) r^2} = e^{-\frac{1}{2}(r/\sigma)^2}
\label{equ:qlim}
\end{equation}
which is the `right half' of a Gaussian bell curve with standard deviation 
\begin{equation}
\sigma=1/\sqrt{2\pi\rho l}. 
\end{equation}
The corresponding distribution of nearest obstacle distances is a Rayleigh distribution
\begin{equation}
p_{l\gg r}(r)= \frac{r}{\sigma^2} e^{-\frac{1}{2}(r/\sigma)^2}.
\label{equ:pnodlim}
\end{equation}
The most probable nearest obstacle distance, i.e. the value of $r$ at which $p(r)$ is maximum, is given by $\sigma$. We note that the accessible volume fraction in Eq.(\ref{equ:qlim}) depends only on the ratio $r/\sigma$. Therefore, all nearest obstacle distance distributions $p_{l\gg r}(r)$ should collapse onto a universal distribution when the distance $r$ is measured in units of $\sigma$. In the long fiber limit, a dense and a dilute Mikado network cannot be distinguished from each other, if the spatial scale is unknown.

\begin{figure}
\includegraphics[width=0.6\linewidth]{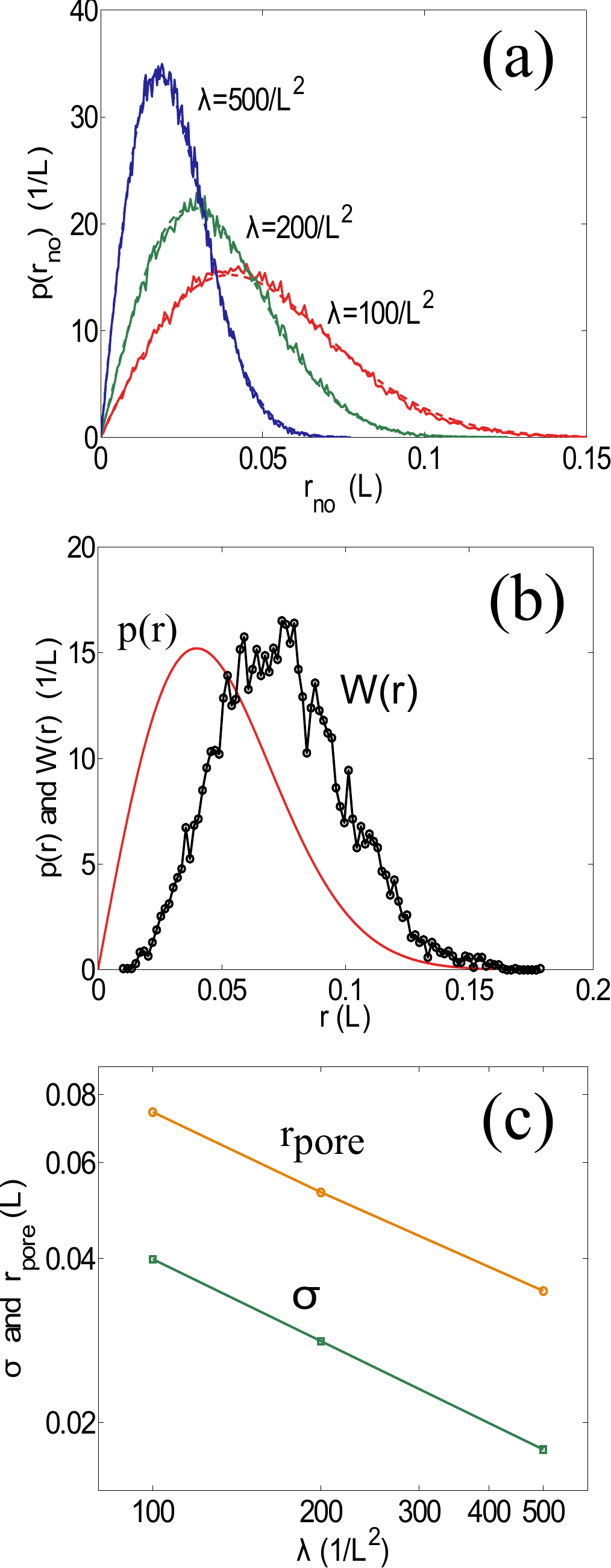}
\caption{\label{fig:fig2} (a) Distribution of nearest obstacle distances in 3D networks of straight line segments, for three different density parameters $\lambda$. Analytical predictions of the Mikado model in the long fiber limit (dashed lines) are compared to numerical simulations (solid lines). The unit of length was set equal to the linear size $L$ of the simulation box, which in turn was equal to the length $l$ of the line segments. (b) Distribution of nearest obstacle distances $p(r)$ (line) and pore size distribution $W(r)$ (line with symbols) in a 3D network of straight line segments, for a density parameter of $\lambda\!=\!100\!/\!L^2$. (c) Most probable obstacle distance $\sigma$ (squares) and average pore size $r_{pore}^{av}$ (circles) as a function of the density parameter $\lambda$. In the long fiber limit, the ratio is constant with $\overline{r}_{pore}/\sigma\approx1.86$.}
\end{figure}

\subsection*{Relating $\sigma$ to line density}
It is remarkable that in the long fiber limit of the Mikado model, the properties of the network are completely determined by the parameter combination $\rho l$, which appears in the quantity $\sigma=1/\sqrt{2\pi\rho l}$.

Remembering the definition of $\rho$ as the volume density of LS centers, we can write
\begin{equation}
\rho l = \frac{N}{V}l = \frac{L_{tot}}{V} =: \lambda,
\end{equation}
where $L_{tot}$ is the total length of all LSs. The new density parameter $\lambda$ corresponds to the {\em total `fiber' length per unit volume}. It follows that
\begin{equation}
\sigma=1/\sqrt{2\pi\lambda}.
\label{equ:sigma}
\end{equation}


\subsection*{Numerical test of the Mikado model \label{sec:numTest}}
In order to test the predictions of the Mikado model, we have simulated random line networks and compared the resulting numerical $p(r_{\rm{no}})$ with the analytical results above. 

In the simulation, each line segment (of constant length $l$) was numerically represented by its center coordinates and a unit direction vector, as depicted schematically in Fig. \ref{fig:fig1}(d). Initially, a list of $N$ such line objects was generated, with the center points distributed randomly throughout a cubic simulation box of linear dimension $L$ (with homogeneous density $\rho=\frac{N}{V}=\frac{N}{L^3}$) and with random, isotropic direction vectors 
\footnote{More precisely, in order to avoid boundary effects, we extended the simulation box on each side by $l/2$ and distributed a correspondingly larger number of $N^* = N (L+l)^3/L^3$ line centers within this extended box.}.

The distribution $p(r_{\rm{no}})$ was determined by randomly choosing $K=10^5$ test points $\vec{R}_{k=1\ldots K}$ within the simulation box, finding the nearest obstacle distance $r_{\rm{no}}(\vec{R}_k)$ for each test point and then computing a histogram of these distances. The distance $r_{\rm{no}}(\vec{R}_k)$ is found by first computing the distances $r_{kn}$ between test point $\vec{R}_k$ and all the lines $n$ of the network and then finding the smallest of those values. Note that the distance $r_{kn}$ between a point and a line segment can be obtained exactly (without any `voxelization' required).

For the numerical test of the Mikado model in the long fiber limit, we prescribed the density parameter $\lambda$, set $L=l=1$ and computed the required number of fibers as $N=\frac{\lambda L^3}{l}$. We found an excellent agreement between the analytical prediction and the simulation (compare Fig. \ref{fig:fig2}(a)).

\subsection*{Relation between the most probable nearest obstacle distance $\sigma$ and the average pore size $\overline{r}_{pore}$}
For any concrete network, it is possible to compute the nearest obstacle distance $r_{\rm{no}}(\vec{R}_0)$ for each spatial point $\vec{R}_0$, resulting in a so-called `Euclidean distance map' (EDM). The pore centers of the network can then be defined as the positions $\vec{R}_0=\vec{R}_{max}^{(i)}$ of the local maxima of the EDM and the pore size distribution $W(r_{pore})$ is the distribution of the distance values $r_{pore}^{(i)} = r_{\rm{no}}(\vec{R}_{max}^{(i)})$ taken at these local maxima.  

Based on our numerically exact simulation of random line networks, as described in Sect.\ref{sec:numTest}, we have computed the pore size statistics $W(r_{pore})$ and compared it to the corresponding distribution $p(r_{\rm{no}})$ of nearest obstacle distances. As expected, $W(r_{pore})$ is peaked at a larger value than $p(r_{\rm{no}})$ (compare Fig. \ref{fig:fig2}(c)). In the long fiber limit, the ratio $\overline{r}_{pore}/\sigma$ between the average pore size and the most probable obstacle distance is a constant, i.e. independent from the density parameter $\lambda$ of the network. This follows from the fact that the distribution $p(r_{\rm{no}})$ is universal in length units of $\sigma$. To demonstrate the constant ratio, we have plotted $\overline{r}_{pore}(\lambda)$ and $\sigma(\lambda)$ double-logarithmically (compare Fig. \ref{fig:fig2}(d)).

\section{Summary}

In this paper we have theoretically investigated random line networks, modelled as isotropic and macroscopically homogeneous distributions of straight line segments in 3D space. In the limiting case when the line segments are much longer than the average poresize $\overline{r}_{pore}$, the distances $r_{\rm{no}}$ of random test points to the nearest line segment are distributed according to a Rayleigh distribution
\begin{equation}
p(r_{\rm{no}})= \frac{r_{\rm{no}}}{\sigma^2} e^{-\frac{1}{2}(r_{\rm{no}}/\sigma)^2}.
\end{equation}
The most probable distance $\sigma$ (peak position of the distribution) is determined by the overall line density $\lambda$, i.e. the total line length per unit volume, by
\begin{equation}
\sigma=\frac{1}{\sqrt{2\pi\lambda}}.
\end{equation}
The average poresize $\overline{r}_{pore}$, defined via the radii of maximum spheres fitting into the pores, is proportional to $\sigma$, with $\overline{r}_{pore} \approx 1.86 \;\sigma$.




\begin{acknowledgments}
This work was supported by grants from Deutsche Forschungsgemeinschaft.
\end{acknowledgments}


\end{document}